\documentclass[dvips]{article}

\usepackage{amsmath,amssymb}
\usepackage{xspace}
\usepackage{cite}
\usepackage{psfrag}
\usepackage{lineno}
\usepackage{icrc2011}

\newcommand{\semhl}{\ensuremath{S_\mathrm{em,\,halo}}}

\newcommand{\semhlm}[1]{\ensuremath{S_\mathrm{em,\,halo}^{\mathrm{\scriptscriptstyle
      #1}}}}
\newcommand{\semprm}[1]{\ensuremath{S_\mathrm{em,\,pure}^{\mathrm{\scriptscriptstyle
      #1}}}}
\newcommand{\semm}[1]{\ensuremath{S_\mathrm{em}^{\mathrm{\scriptscriptstyle
      #1}}}}
\newcommand{\smum}[1]{\ensuremath{S_\mu^{\mathrm{\scriptscriptstyle
      #1}}}}
\newcommand{\smillam}[1]{\ensuremath{S_{1000}^{\mathrm{\scriptscriptstyle
      #1}}}}
\newcommand{\sigratm}[1]{\ensuremath{S_\mu^{\mathrm{\scriptscriptstyle
      #1}}/S_\mathrm{em}^{\mathrm{\scriptscriptstyle
      #1}}}}

\newcommand{\gss}{0.7}

\newcommand{\erange}[2]{\ensuremath{10^{#1}-10^{#2}}~eV}
\newcommand{\arange}[2]{\ensuremath{#1^\circ-#2^\circ}}
\newcommand{\xmax}{\ensuremath{X_\mathrm{max}}}
\newcommand{\xmaxv}{\ensuremath{X_\mathrm{max}^\mathrm{v}}}
\newcommand{\gsm}{g/cm${}^2$}

\newcommand{\sigrat}{\ensuremath{S_\mu/S_\mathrm{em}}}
\newcommand{\sem}{\ensuremath{S_\mathrm{em}}}
\newcommand{\smu}{\ensuremath{S_\mu}}

\newcommand{\logen}{\lg(E/\mathrm{eV})=}

\psfrag{Xmax}[rb][rb][\gss]{\xmaxv, \gsm}
\psfrag{Xmaxv}[rb][rb][\gss]{\xmaxv, \gsm}
\psfrag{SmuSem}[rc][rc][\gss]{$\mathbf{\sigrat}$}
\psfrag{HaloFraction}[rb][rc][\gss]{$\mathbf{\semhl/\sem}$}
\psfrag{SemHaloSmu}[rb][rc][\gss]{$\mathbf{\semhl/\smu}$}
\psfrag{SemHaloFraction}[r][r][\gss]{$\mathbf{\semhl/\sem}$}
\psfrag{Zth*180/TMath::Pi\(\)}[r][r][\gss]{\bf{Zenith angle}}
\psfrag{Zenith angle}[r][r][\gss]{\bf{Zenith angle}}
\psfrag{Smu[3][3]/(Sem[3][3]+Sgm[3][3])}[rc][rc][\gss]{$\mathbf{\sigrat}$}
\psfrag{Smu[3][3]/\(Sem[3][3]+Sgm[3][3]\)/pow\(cos\(Zth\),\2551.2\)}[rb][rb][\gss]{$\sigrat\cos^{1.2}(\theta)$}

\title{Precise determination of muon and electromagnetic shower contents from shower universality property}

\newcommand{\etal}{\MakeLowercase{\textit{et al. }}} 
\shorttitle{A.~Yushkov \etal Precise determination of muon and EM shower contents}

\authors{A.~Yushkov$^{1}$, M.~Ambrosio$^{2}$, C.~Aramo$^{2}$, D.~D'Urso$^{2}$, F.~Guarino$^{2,3}$, L.~Valore$^{2}$}
\afiliations{$^1$Universidade de Santiago de Compostela, Santiago de
  Compostela, Espa\~na 15782\\
$^2$INFN Sezione di Napoli, via Cintia, Napoli, Italia 80125\\ 
$^3$Universit\`{a} di Napoli ``Federico~II'', via Cintia, Napoli, Italia 80125\\
}
\email{yushkov.alexey@gmail.com}

\abstract{We present two new aspects of Extensive Air Shower (EAS)
  development universality allowing to make accurate estimation of
  muon and electromagnetic (EM) shower contents in two independent
  ways. In the first case, to get muon (or EM) signal in water
  Cherenkov detectors it is enough to know the vertical depth of shower
  maximum and the total signal. In the second case, the EM signal can
  be calculated from the primary particle energy and the zenith
  angle. In both cases the parameterizations of muon and EM signals
  are almost independent on primary particle nature, energy and zenith
  angle.} 

\begin{document}
\maketitle

\begin{figure}[tbh]
\centering\includegraphics[width=0.46\textwidth]{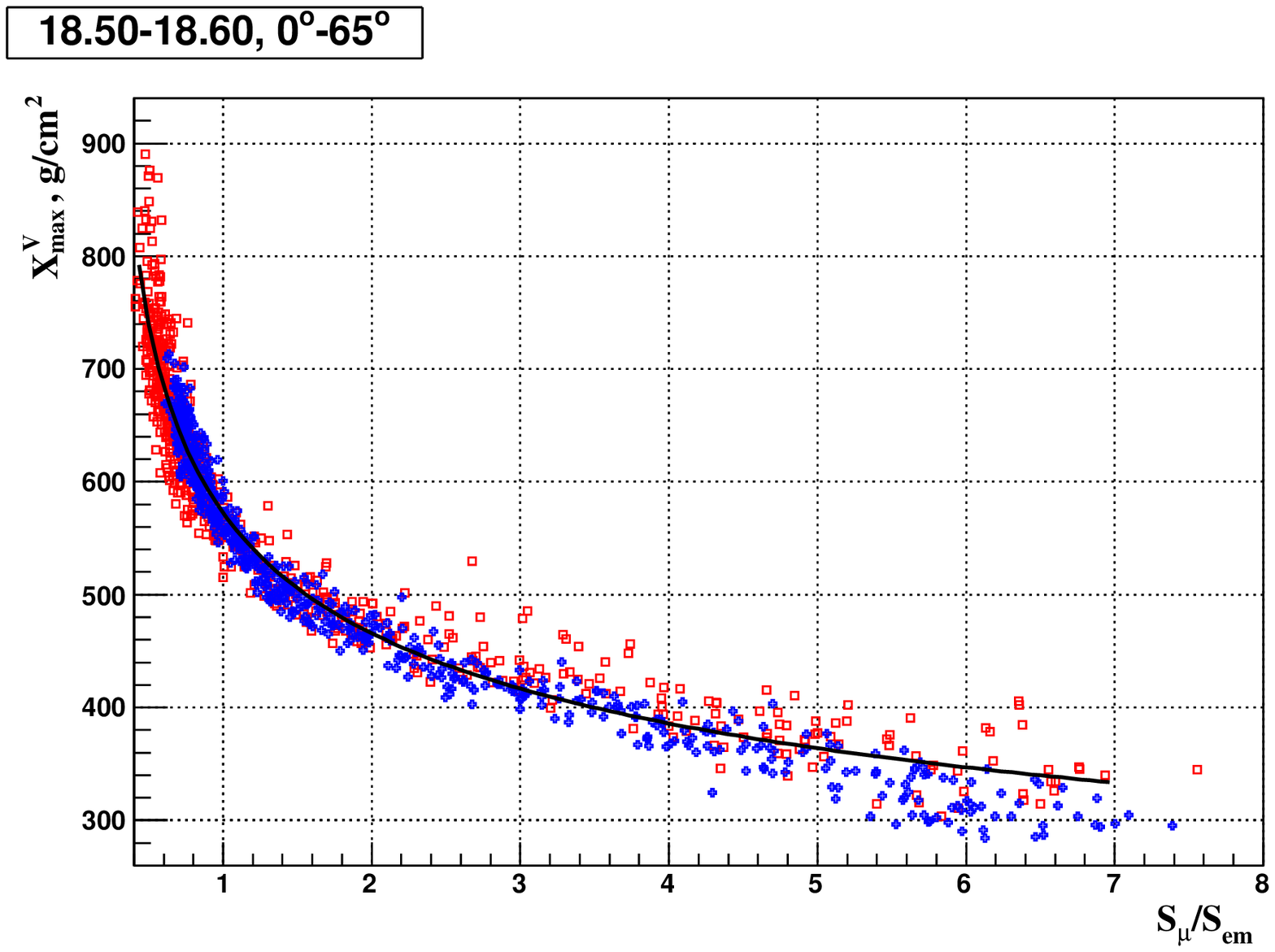}
\centering\includegraphics[width=0.46\textwidth]{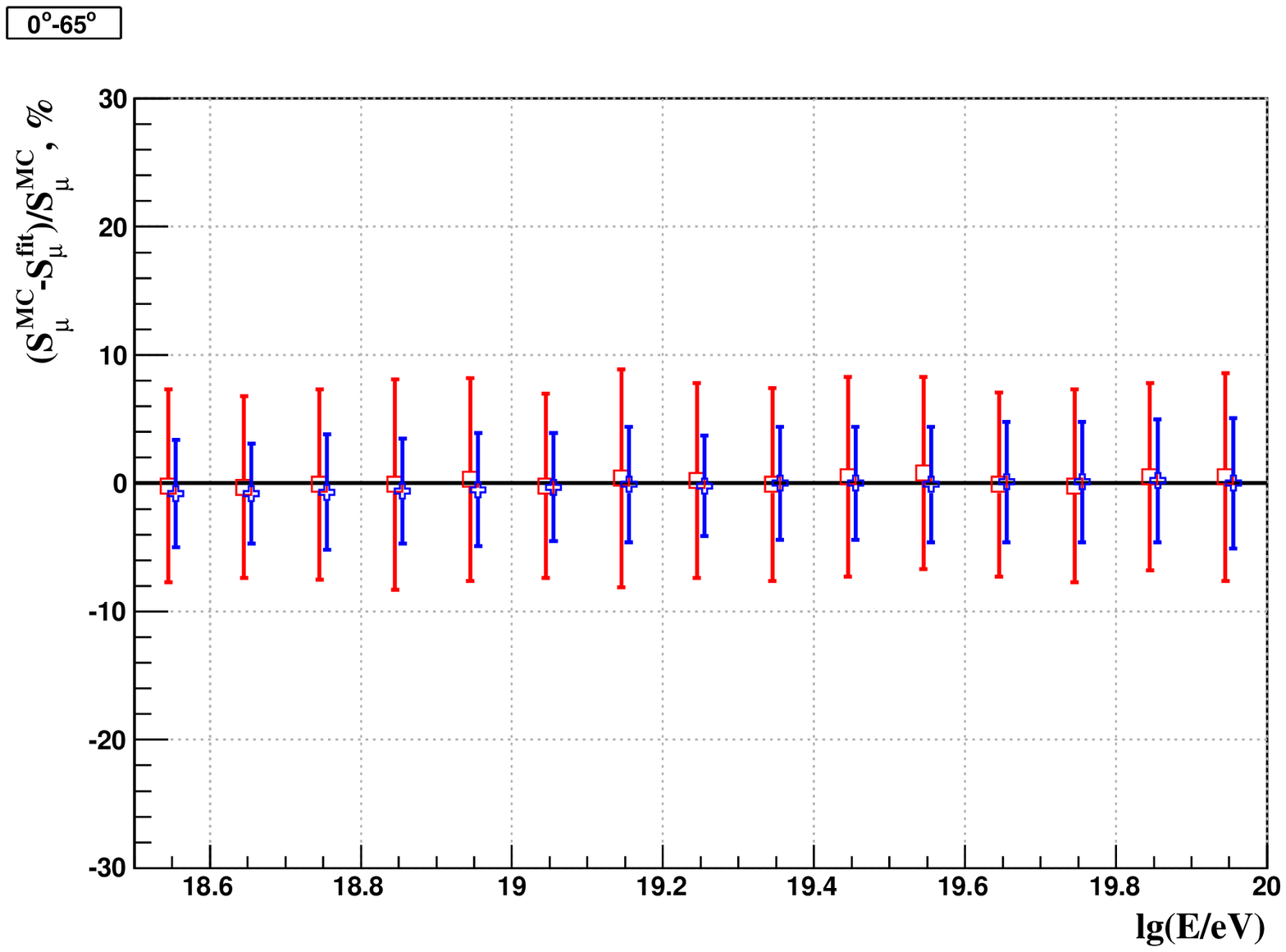}
\caption{Top: ratio of ground plane signals at 1000~m in water
  Cherenkov detectors \sigrat\ vs vertical depth of shower maximum
  \xmaxv\ in \erange{18.5}{18.6} energy range for QGSJET~II. Black line is the fit
  in the form~(\ref{eq:fit}).  Bottom: means and RMS of distributions
  of relative difference between MC simulated muon signals
  $S_\mu^\mathrm{MC}$ and muon signals derived from the
  fit~(\ref{eq:fit}) $S_\mu^\mathrm{fit}$, calculated with the unique
  set of parameters for all energy bins: $A=538$, $b=-0.25$,
  $a=-0.22$. Protons~---~red squares, iron~---~blue crosses.}
\label{muemxmax}
\end{figure}

\begin{figure}[tbh]
\centering\includegraphics[width=0.46\textwidth]{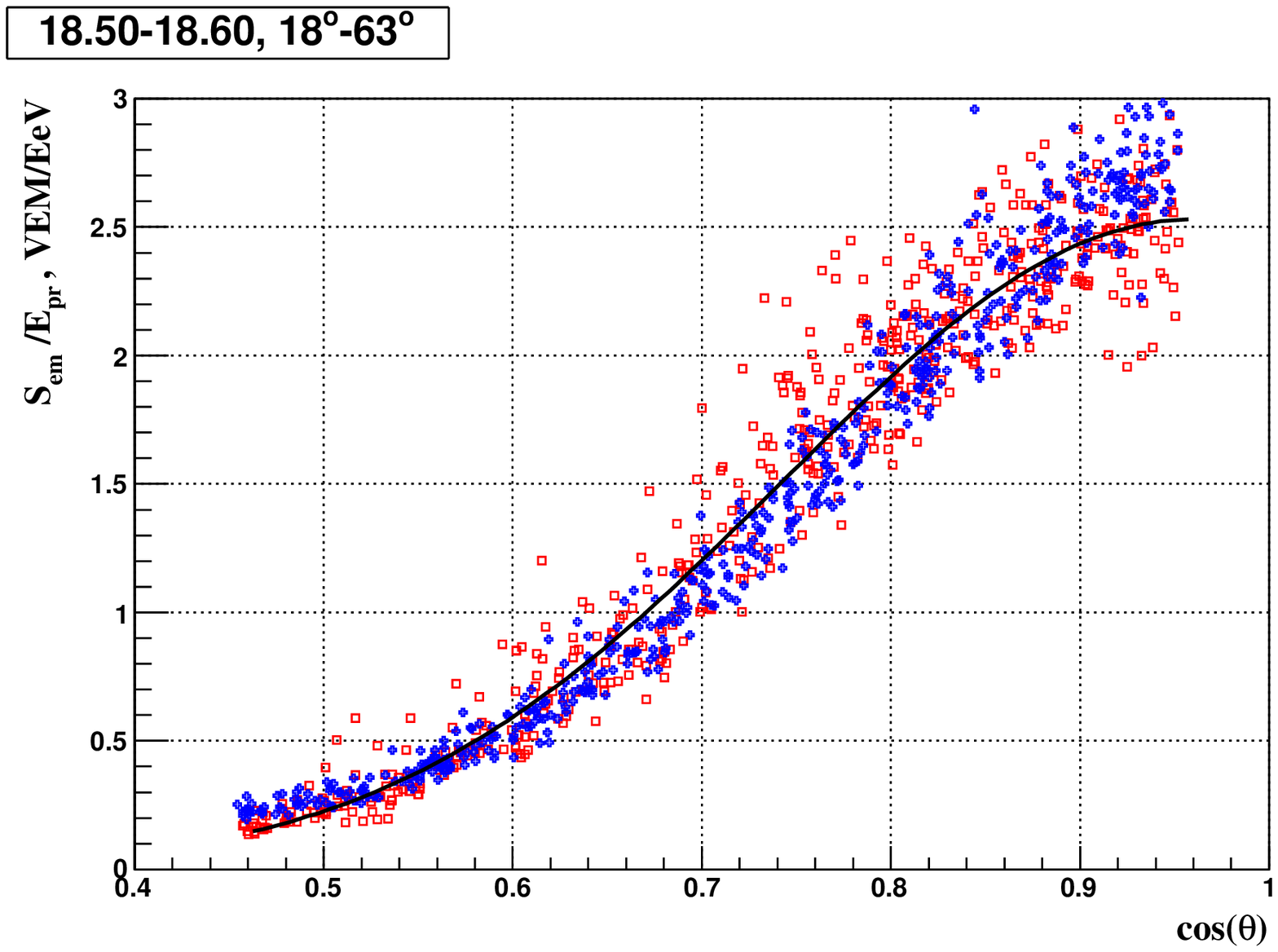}
\centering\includegraphics[width=0.46\textwidth]{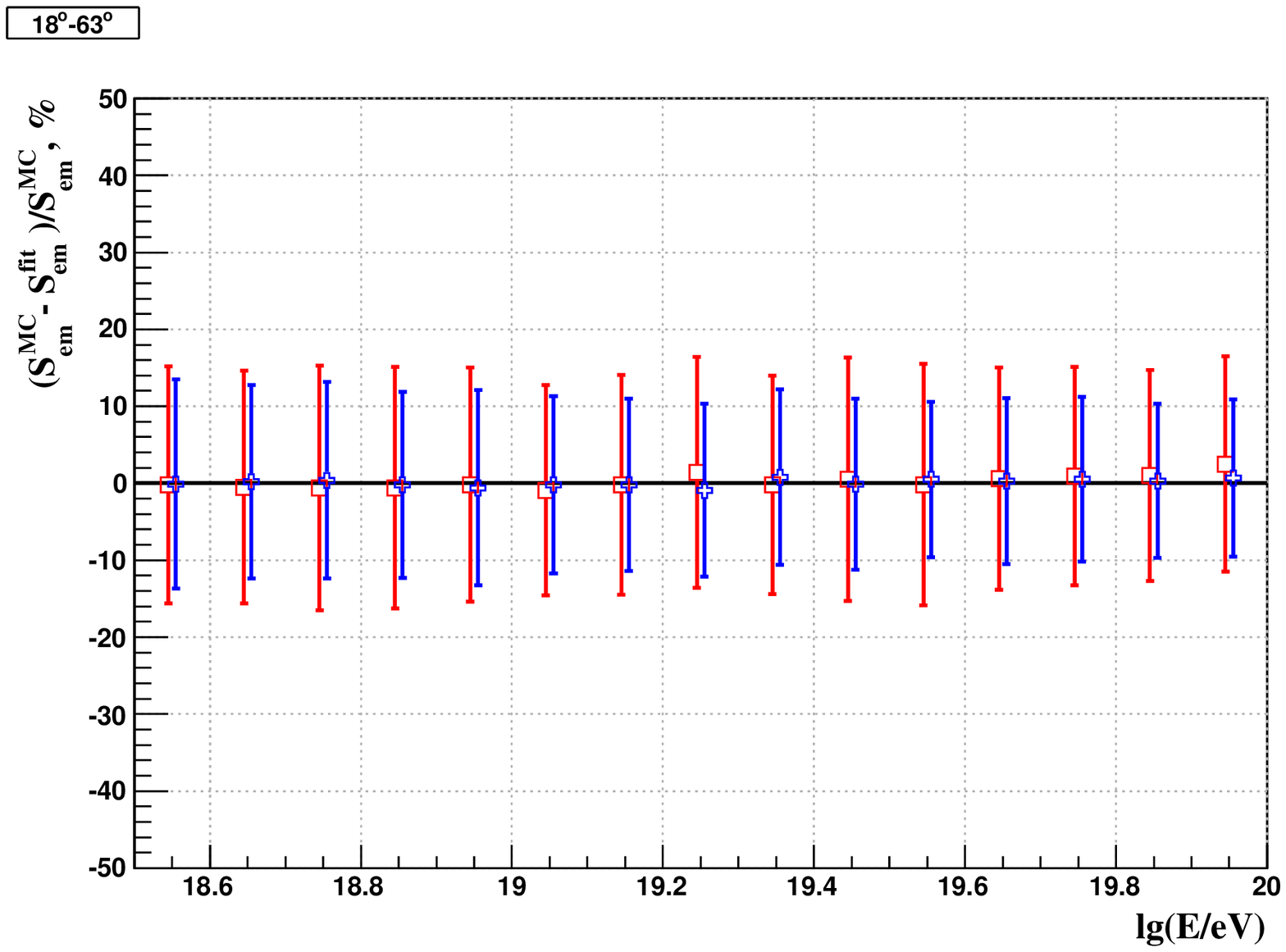} 
\caption{Top: EM ground signals at 1000~m in water Cherenkov detectors vs
  $\cos(\theta)$ in \erange{18.5}{18.6} energy range and
  $\theta=18^\circ-63^\circ$ zenith angle range for QGSJET~II. Black line is the fit in
  the form~(\ref{eq:semcos}). Bottom: means and RMS of distributions
  of relative difference between MC simulated EM signals
  $\sem^\mathrm{MC}$ and EM signals derived from the
  fit~(\ref{eq:semcos}) $\sem^\mathrm{fit}$, calculated with the
  unique set of parameters for all energy bins: $\sem^0=2.53$,
  $c_0=-3$, $c_1=0.96$, $\lambda=0.012$. Protons~---~red squares,
  iron~---~blue crosses.}
\label{SemcosE}
\end{figure}

\begin{figure}[tbh]
\centering\includegraphics[width=0.46\textwidth]{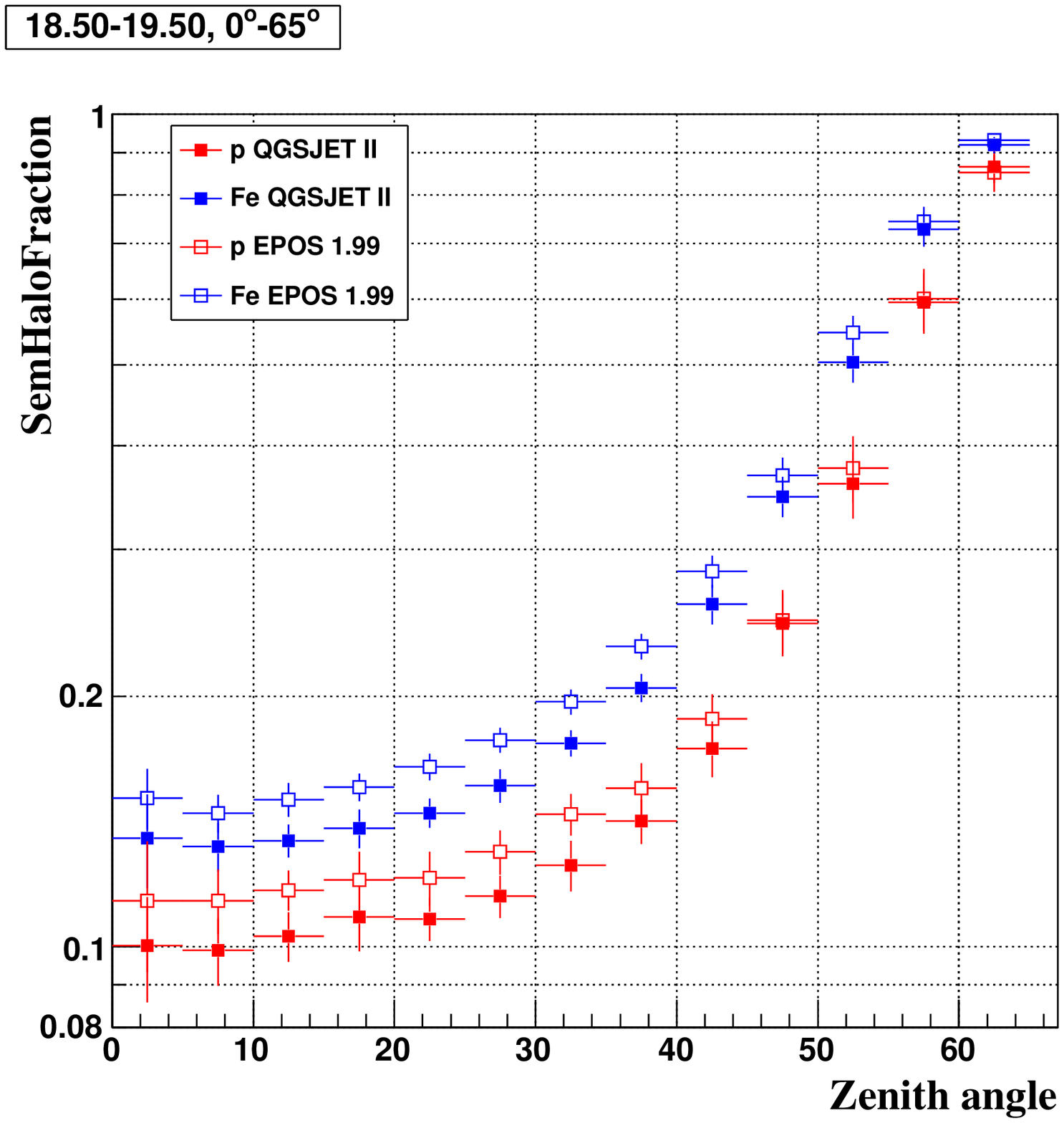}
\centering\includegraphics[width=0.46\textwidth]{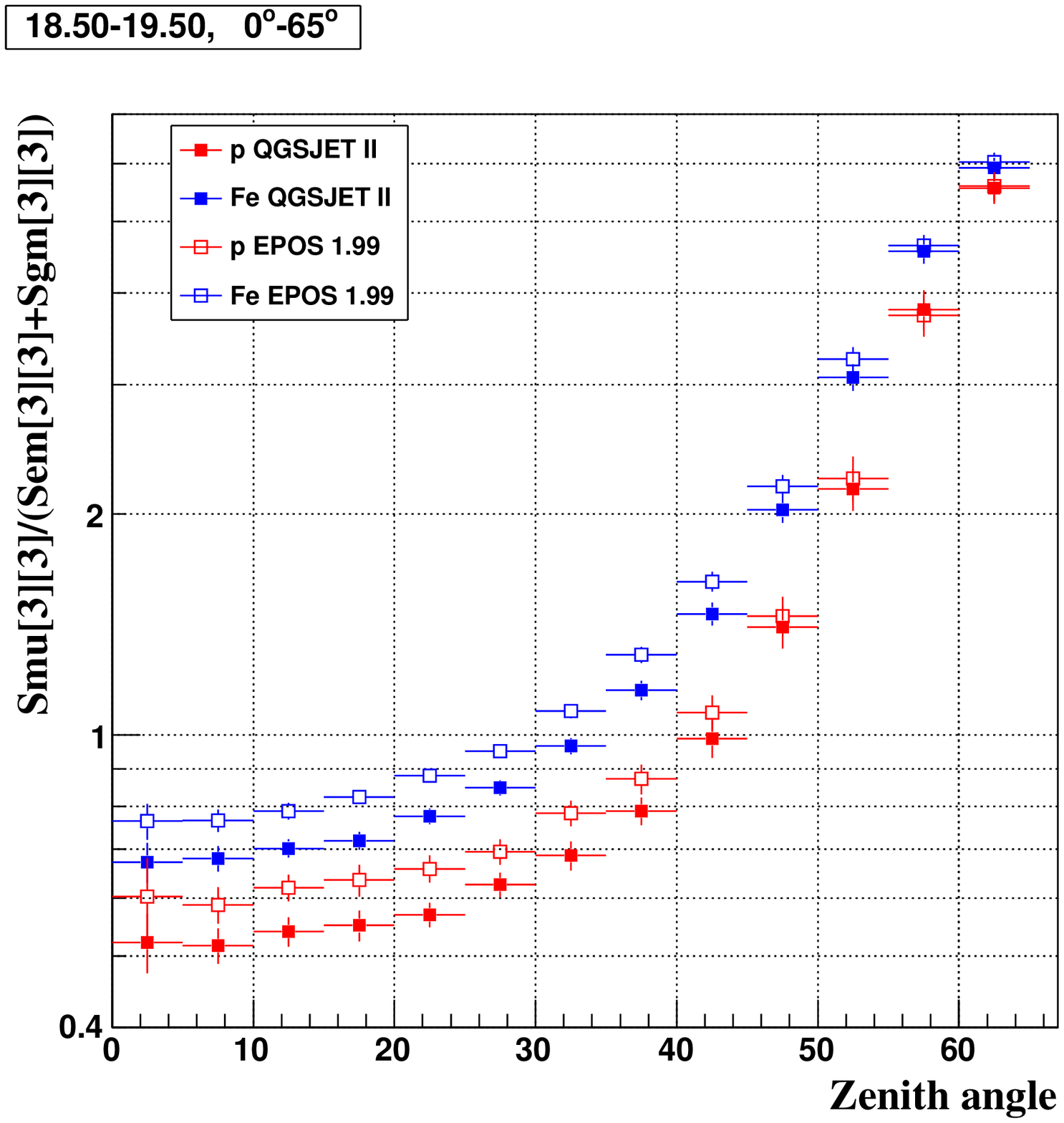}
\caption{Top: EM muon halo fraction \semhl\ of total EM signal
  \sem\ vs zenith angle. Bottom: \sigrat\ dependence on the zenith
  angle. $\logen18.50-19.50$.}
\label{fig:vszenith}
\end{figure}

\section*{Introduction}
Mass composition of ultra-high-energy cosmic rays (UHECR) can be
studied only indirectly with large EAS arrays. The contemporary
measurement of longitudinal and lateral shower characteristics in
hybrid experiments like the Pierre Auger
Observatory~\cite{PAO_proto_NIMA2004} provides the possibility to
combine several primary mass sensitive EAS parameters (such as depth
of shower maximum and muon shower content) to achieve the best primary
particle mass discrimination. Unfortunately, the lack of reliable
information on hadronic interaction properties at these energies
causes large uncertainties in the simulations of EAS characteristics
and in turn brings large uncertainties in mass composition analysis
results (see e.g. recent review~\cite{bluemer_crknee2009}).

In this paper we propose two simple, independent and accurate methods
to determine muon and EM shower contents in hybrid experiments and
briefly discuss a possible way to test and adjust interaction models
properties in a primary mass independent way. We also hope that the
proposed EAS-universality-based correction of the interaction models
will allow to perform mass composition analysis with the use muon EAS
content in less interaction model dependent manner.

The present study is performed making use of around 50000 showers,
generated with
CORSIKA~6.735~\cite{corsika}/QGSJET~II~\cite{qgsjetiia}/Fluka~\cite{fluka1}
(see~\cite{ya_PRD2010} for full list of references and more details)
and CORSIKA~6.900/EPOS~1.99~\cite{werner_epos2006}/Fluka for $E^{-1}$
spectrum in the energy range $\logen18.5-20.0$ and uniformly
distributed in $\cos^2{\theta}$ in zenith angle interval
$\theta=0^\circ-65^\circ$. EM component thinning was set to $10^{-6}$,
the observation level was at 870~\gsm, geomagnetic field was set to
the value of the site of the Auger Observatory in Malarg\"ue.
The expected signal $S$ in Cherenkov Auger-like detectors was
calculated according to the sampling procedure described
in~\cite{billoir_sampl_2008,ave_munum_2007} with the use of the same
GEANT~4 lookup tables as in~\cite{ave_munum_2007}. Differently
from~\cite{ave_munum_2007} in this work the muon signal $S_\mu$
includes only signal from muons crossing the Cherenkov detector, while
signal from EM particles, originating from muon decays, is included in
the EM signal.

\section{Showers at the same $\mathbf{\xmaxv}$}

Of all aspects of universality of shower development, we will be
interested only in dependence of EM and muon signals on the distance
of shower maximum to the ground and on the zenith angle. Let's
consider the Auger-like experimental setup~\cite{PAO_proto_NIMA2004}
and ground-plane signal in water Cherenkov detectors at 1000 meters
from the shower core.

Comparing shower characteristics dependence on the vertical depth of
shower maximum \xmaxv\ one finds a very interesting property. Clearly,
to have the same \xmaxv\ an average proton shower has to be more
inclined than the iron shower of the same energy. Therefore, the EM
component in the proton shower will attenuate more while reaching the
ground from the shower maximum and it turns out that
$S_\mathrm{em}^\mathrm{Fe}/S_\mathrm{em}^\mathrm{p}$ ratio becomes
almost equal to the $S_\mu^\mathrm{Fe}/S_\mu^\mathrm{p}$ one, that
allows to state a new shower universality property: the ratio of the
muon signal to the EM signal \sigrat\ is the same for all showers,
reaching the maximum at the same vertical depth \xmaxv, independently
on the primary particle nature, primary energy and incident zenith
angle (for the energy and angular ranges considered here). This
property is illustrated in Fig.~\ref{muemxmax}, where the dependence
of \sigrat\ on \xmaxv\ for $p$ and Fe primaries is shown in
$\logen18.5-18.6$ energy bin. The functional dependence between
\xmaxv\ and \sigrat\ turns out to be very simple and quasi-universal
for all energies and primaries. The following function
\begin{equation}
\label{eq:fit}
\xmaxv=A(S_\mu/S_\mathrm{em}+a)^b 
\end{equation}
has been used to fit the data in 15 energy bins
$\Delta\lg(E/\mathrm{eV})$=0.1 and the fit parameters have been
found to be stable across the entire energy range. Using the
functional dependence of \sigrat\ on \xmaxv\ and
$S_\mathrm{1000}=S_\mathrm{em}+S_\mu$ one easily gets the equation,
which allows to obtain the muon signal from vertical depth of maximum
and total signal in water Cherenkov detectors:
\begin{equation}
\label{eq:mufit}
S_\mu^\mathrm{fit}=\frac{S_\mathrm{1000}}{1+1/(\left(\xmaxv/A\right)^{1/b}-a)}.
\end{equation}
We calculated the difference between the Monte-Carlo (MC) simulated
muon signal $S_\mu^\mathrm{MC}$ and the muon signal obtained from the
fit $S_\mu^\mathrm{fit}$. In Fig.~\ref{muemxmax} we plot the behaviour
of the mean and RMS values of these distributions for various
energies, obtained with the unique set of fit parameters $A=538$,
$b=-0.25$ and $a=-0.22$, representing the averages over 15
$\Delta\lg(E/\mathrm{eV})$=0.1 energy bins. It is seen that the
estimates of muon signals are unbiased with less than 1\% deviation of
the mean reconstructed muon signal from the MC one for all primaries
and the RMS values are small: 8\% for protons and around 5\% for
oxygen and iron (though we don't show results for oxygen, we use
oxygen showers together with proton and iron ones to perform fits).

In case of EPOS~1.99 the same universality holds and the fit in the
form~(\ref{eq:fit}) also provides good description of the simulated
data, but, as expected, the coefficients of the fit are different from
those for QGSJET~II.

\section{Showers at the same zenith angles}
Another universality property follows from the study of showers
arriving at the same zenith angles. In this case the average iron
shower has to cross larger slant distance from \xmax\ to the ground
with respect to the average proton shower and this almost equalizes EM
signals for both primaries at the observation level in a wide range
of zenith angles. For the signal at 1000 meters in the Cherenkov water
detectors notable discrepancies between $p$ and Fe EM showers components
are observed for nearly vertical showers
($\theta<18^{\circ},\ \cos^2(\theta)>0.9$) and very inclined ones
($\theta>63^{\circ},\ \cos^2(\theta)<0.2$). In the first case the path
from \xmax\ to the ground for $p$ and Fe showers is almost the same. For
inclined showers the difference is caused by the EM halo from muon
decays and larger number of muons in iron showers brings to a larger
EM halo signal.

Looking at the showers at different zenith angles one samples
longitudinal showers profiles, for this reason it is natural to try to
describe the dependence of the EM signal on $\cos(\theta)$ with
Gaisser-Hillas type function, using $\cos(\theta)$ as variable instead
of \xmax:
\begin{multline}
\label{eq:semcos}
\frac{\sem(E,\,\theta)}{E}\left[\frac{\mathrm{VEM}}{\mathrm{EeV}}\right]=\sem^0\left(\frac{\cos(\theta)-c_0}{c_1-c_0}\right)^\alpha\times\\ \times\exp\left(\frac{c_1-\cos(\theta)}{\lambda}\right),
\end{multline}
where $\alpha=(c_1-c_0)/\lambda$; $\sem^0$ (signal at maximum), $c_0$,
$c_1$ (cosine of angle at which \sem=$\sem^0$) and $\lambda$ are fit
parameters. The fit parameters $\sem^0$ and $c_1$ change by less than
10\% and 3\% correspondingly across the entire range of energies (when
one makes fits in 15 energy bins $\Delta\lg(E/\mathrm{eV})$=0.1 from
$\logen18.5$ to $\logen20.0$), while $c_0$ changes quite chaotically
from $0$ to $-20$ (this causes $\lambda$ to change also). We have
found that fixing $c_0$ (similarly to~\cite{ave_munum_2007}) to any
negative value within this range, we obtain a good universal fit and
$\lambda$ changes in this case by less than 15\%. Finally, we used the
following average values (except for $c_0$ that was fixed to $-3$) of
the coefficients $\sem^0=2.53$, $c_0=-3$, $c_1=0.96$,
$\lambda=0.012$. The results of the fit and the difference between the
MC simulated EM signal $\sem^\mathrm{MC}$ and the EM signal obtained
from the fit $\sem^\mathrm{fit}$ are shown in Fig.~\ref{SemcosE}. The
accuracy of the EM signal reproduction for all energy bins is such
that one gets an unbiased estimate of \sem\ with RMS below 15\% for
proton and 13\% for iron showers.

Our calculations demonstrate that the universality of EM signal
dependence on zenith angle holds true also in case of EPOS~1.99.

\section{$\mathbf{\sigrat}$ universality in respect to interaction
  models for $\mathbf{\theta>45^\circ}$}

Phenomenologically the angular region \arange{45}{65} is of interest
since with increase of the zenith angle the EM component produced
mostly in $\pi^0$ decays at the initial EAS development stages is
largely absorbed in the atmosphere and EM halo from muon decays starts
to play a remarkable role (Fig.~\ref{fig:vszenith}). One expects in
this case that the behavior of the \sigrat\ ratio should become less
sensitive to the properties of the interaction models since with
increase of the angle it more and more reflects the equilibrium state
between muons and EM halo from muons decays and interactions. To
illustrate quantitatively this process let us write the \sigrat\ ratio
for QGSJET~II as
$$\sigratm{QGS}=\frac{\smum{QGS}}{\semhlm{QGS}+\semprm{QGS}},$$ here
\semhlm{QGS} is the EM halo signal from muons, \semprm{QGS} is EM
signal from everything else except muons. Then for EPOS~1.99 one gets
$$\sigratm{EPOS}=\frac{\mu\smum{QGS}}{\mu\semhlm{QGS}+\varepsilon\semprm{QGS}},$$
where $\mu=\smum{EPOS}/\smum{QGS}$ and
$\varepsilon=\semprm{EPOS}/\semprm{QGS}$ are the scaling factors
between muon and EM signals of the models and we have taken into account
that $\semhl\propto\smu$ and so $\mu=\semhlm{EPOS}/\semhlm{QGS}$.
In these notations one gets
\begin{equation}
\label{eq:vszenith}
\frac{\sigratm{EPOS}}{\sigratm{QGS}}=1+\frac{\mu-\varepsilon}{\varepsilon+\mu(\semhlm{QGS}/\semprm{QGS})}.
\end{equation}
One can see from Eq.~(\ref{eq:vszenith}) that with the increase of the
zenith angle and hence of \semhlm{QGS}/\semprm{QGS} the difference between
models in \sigrat\ is decreasing as shown in the bottom panel of
Fig.~\ref{fig:vszenith}. Let us note that from the approximate
equality of \sigrat\ ratios for different models it follows that for
any primary nucleus ($p$, O, Fe etc.) the following equality holds
\begin{equation}
\label{eq:allscales}
\frac{\smillam{EPOS}}{\smillam{QGS}}\approx\frac{\smum{EPOS}}{\smum{QGS}}\approx\frac{\semm{EPOS}}{\semm{QGS}}.
\end{equation}
This, in turn, means that Eq.~(\ref{eq:mufit}) in this angular range
provides an almost model-independent estimate of the muon signal. In
fact, since the muon signal scales in the same way as the total
signal, if one applies e.g. Eq.~(\ref{eq:mufit}) with fit parameters
for QGSJET~II to the data simulated with EPOS~1.99, the total signal
of EPOS~1.99 will give correct normalization for the muon signal. 
The difference between models in \xmaxv\ and in the functional
dependence on \xmaxv\ will play only a minor role. As it will be
demonstrated elsewhere~\cite{mimmo_icrc2011} the muon signal for
EPOS~1.99 can be retrieved with the use of the QGSJET~II fit
parameters with accuracy of 3--5\%.

\section*{Conclusions}
We have presented two new EAS universality properties providing two
independent ways to access EM and muon shower contents. We have shown
that these properties can be described with simple parametrizations
which are valid in wide energy and zenith angle ranges, and are
independent on the primary particle nature. We believe that these
universality properties can be used in hybrid experiments for mass
composition studies, for primary and missing energy estimates and for
tests of hadronic interaction models. One of the possible strategies
lies in the simultaneous application of both universality properties
to the data. It is clear that parametrizations~(\ref{eq:mufit})
and~(\ref{eq:semcos}) will give consistent estimates of muon and EM
shower contents only in case of correct description of the hadronic
interaction properties by the particular model. Another interesting
strategy can be pursued in hybrid experiments equipped with muon
detectors. For zenith angles above 45 degrees where the EM halo plays
an important role, this universality property can be used for the
determination of the depth of the shower maximum in almost interaction
model independent way taking advantage of 100\% ground array duty
cycle with respect to 10\% one of the fluorescence telescopes. On the
other hand for angles below 45 degrees the difference in behaviour of
\sigrat\ between models should be large enough so that with
simultaneous knowledge of \smu, \sem\ and \xmax\ one could be able to
check the predictions of hadronic models quite easily
using~(\ref{eq:fit}) and comparing e.g. the parameterized \xmax\ with
the measured one.

Finally, we would like to dwell on the problem of muon excess in the
real data compared to predictions of the interaction
models~\cite{ave_munum_2007,abuzayyad_mass2000,engel_icrc2007}. Since
the muon content of EAS is highly model-dependent and the UHECR mass
composition is still unknown, this muon excess can be expressed only
in terms of a relative excess with respect to the prediction of a
given hadronic interaction model for a given primary like (real
signal)/(MC signal for protons). In~\cite{mimmo_icrc2011} we show that
for the zenith angles above 45 degrees it is possible to get muon
shower content from real data in almost interaction model independent
way (see application to the data of the Pierre Auger Observatory
in~\cite{auger_nmu_icrc2011}). This, in turn, provides us with EM
shower content which is weakly sensitive to the mass of the primary
particle and hence allows to find the absolute scaling factor (real
signal protons)/(MC signal protons). Performing such scaling on the
one hand will diminish the difference in predicted muon signals
between different interaction models, and on the other hand will open
the possibility to use the muon signal for mass composition studies.

\subsubsection*{Acknowledgements}
We are very grateful to Maximo Ave and Fabian Schmidt for kind
permission to use their GEANT~4 lookup tables in our calculations of
signal from different particles in Auger water Cherenkov detectors.


\end{document}